\pdfoutput=1
\documentclass[11pt]{article}

\usepackage[T1]{fontenc}
\usepackage[utf8]{inputenc}
\usepackage{lmodern}
\usepackage{microtype}

\usepackage[margin=1in]{geometry}

\usepackage{amsmath,amssymb,amsfonts}

\usepackage{graphicx}
\usepackage{float}

\usepackage{cite}
\usepackage{hyperref}
\hypersetup{
  colorlinks=true,
  linkcolor=blue,
  citecolor=blue,
  urlcolor=blue
}

\IfFileExists{braket.sty}{\usepackage{braket}}{%
  \newcommand{\bra}[1]{\left\langle #1\right|}
  \newcommand{\ket}[1]{\left|#1\right\rangle}

}

\newcommand{\maybeincludegraphics}[2][]{%
  \IfFileExists{#2.pdf}{\includegraphics[#1]{#2}}{%
    \IfFileExists{#2.png}{\includegraphics[#1]{#2}}{%
      \IfFileExists{#2.jpg}{\includegraphics[#1]{#2}}{%
        \fbox{\parbox{0.9\linewidth}{\centering Missing figure file: \texttt{#2}.*}}%
      }%
    }%
  }%
}

\graphicspath{{./}{figures/}{Figures/}}

\title{Population--Coherence Routes to Purity in Page-Type Models of Black-Hole Evaporation}
\author{Jos\'e J. Gil\\
Independent Researcher, 29690 Casares Costa, Spain\\
\texttt{ppgil@unizar.es}}
\date{\today}

\begin{document}
\maketitle

\begin{abstract}
We revisit the black-hole information problem from the viewpoint of a population--coherence decomposition of density-matrix purity. Building on a previously developed formalism for $n$-dimensional density matrices, we characterize each state by a normalized global purity index and two complementary indices, which quantify the contributions of level populations and coherences. This yields a simple quadratic relation and a geometric representation in a ``population--coherence plane'', where different routes to purity can be distinguished. In the two-level case we construct explicit families of states with identical spectra and global purity but opposite internal structure, realizing population-dominated and coherence-dominated routes. We then apply this framework to a standard Page-type evaporation model without an explicit Hamiltonian, in which a black hole and its Hawking radiation form a bipartite pure state with varying Hilbert-space dimensions. Using known results for typical reduced states in large dimensions, we analyze the behavior of population and coherence components of purity along the evaporation process. Under the physically motivated requirement that, in this energy-free setting, the radiation populations remain nearly uniform in the chosen basis, we show that the late-time recovery of purity must be coherence-dominated: the global purity of the radiation approaches unity while the population index stays small and the coherence index carries essentially all the purity.
\end{abstract}


\section{Introduction}
\label{sec:intro}

The black-hole information problem remains one of the central conceptual challenges at the interface between quantum mechanics and gravitation. In Hawking's semiclassical picture, an initially pure state collapsing to form a black hole evolves into a mixed state of nearly thermal radiation, apparently violating unitarity~\cite{Hawking1975}. The tension arises because the standard quantum description of closed systems demands that the global evolution be unitary, preserving the purity of the global state, whereas the effective description of Hawking radiation suggests an irreversible loss of information.

Over the last decades, a variety of proposals have been put forward to reconcile these two viewpoints. Some approaches modify the semiclassical dynamics or the near-horizon structure of spacetime; others exploit nonlocal correlations or invoke subtle aspects of quantum gravity. More recently, the development of the ``island formula'' and quantum extremal surfaces has provided a framework in which the entanglement entropy of Hawking radiation can exhibit the Page-curve behavior expected from unitary evaporation models~\cite{Page1993,Almheiri2019,Almheiri2020}. In these accounts, the fine-grained entanglement entropy of the radiation initially grows, reaches a maximum near the Page time, and subsequently decreases back to zero as the black hole evaporates, restoring the purity of the global state.

Most discussions of the information problem, including those based on Page curves and island constructions, focus on entropic quantities as the main diagnostic of information flow. This is natural, given that the paradox is often phrased in terms of entropy increase and decrease. However, from the point of view of quantum statistical description, a density matrix $\rho$ contains more structure than a single entropic number. Even at the level of second-order statistics, the spectrum of $\rho$ and its matrix structure encode not only how ``pure'' the state is, but also where that purity resides: in the level populations (diagonal entries) or in the coherences (off-diagonal correlations) in a given physically meaningful \emph{intrinsic} basis for which the real part of the density matrix is diagonal, while the off-diagonal elements are pure imaginary.

In a different context, namely classical polarization optics and general density-matrix analysis, previous work has shown that it is often useful to decompose the global purity of a state into distinct contributions associated with populations and coherences~\cite{GilSourcesAsymmetry}. In that work, a normalized global purity index $P_{(n)}$ for an $n$-dimensional density matrix was complemented by two auxiliary indices $P_p$ and $P_c$, capturing, respectively, the asymmetry of the intrinsic populations and the asymmetry due to intrinsic coherences (or correlations). For general $n$, these indices obey a relation of the form
\begin{equation}
  P_{(n)}^2 = P_p^2 + \alpha_n P_c^2,
  \qquad
  \alpha_n = \frac{n}{2(n-1)},
  \label{eq:Pn2_Pp_Pc_intro}
\end{equation}
which, in the two-dimensional case, reduces to the simple Euclidean decomposition
\begin{equation}
  P_{(2)}^2 = P_p^2 + P_c^2,
\end{equation}
Geometrically, each state can be represented by a point in the $(P_p,P_c)$ plane, with the global purity $P_{(2)}$ playing the role of the radial coordinate and the relative weights of $P_p$ and $P_c$ determining the angular coordinate. This ``population--coherence plane'', which is  easily generalized to $n$ dimensions, offers a compact way of visualizing how purity is distributed within a density matrix.

The present work explores how this population--coherence decomposition of purity can sharpen our description of the black-hole information problem. Instead of asking only whether the entropy of Hawking radiation increases or decreases with time, we ask two related questions. \textit{First}, when the radiation becomes purer at late times, is that purity stored primarily in the intrinsic level populations, or in intrinsic coherences? \textit{Second}, to what extent can a state be locally thermal in its energy populations while still being globally pure?

These questions become particularly pressing in the context of unitary evaporation scenarios. If we require that the global evolution be unitary, then the late-time radiation must be nearly pure. At the same time, if we insist that local measurements in the energy basis see a spectrum that remains close to thermal throughout the evaporation process, then the diagonal of the radiation density matrix should not deviate strongly from a thermal distribution. Under these conditions, it is natural to suspect that any recovery of information must manifest in the off-diagonal structure of the density matrix, that is, in the intrinsic coherences.

The population--coherence framework provides a precise language to express this intuition. In particular, we distinguish two limiting cases: a \emph{population-dominated route to purity}, in which $P_p$ carries most of the global purity $P_{(n)}$ and the density matrix is close to diagonal in the chosen basis, so that information is encoded mainly in anisotropic level populations leading to significant deviations from a thermal spectrum; and a \emph{coherence-dominated route to purity}, in which $P_c$ carries most of $P_{(n)}$ while $P_p$ remains small, so that the populations remain nearly thermal and the information is stored in nontrivial coherences and correlations that are invisible to measurements probing only the diagonal of $\rho$.

Our goal in this paper is to show that, within a standard Page-type model of evaporation, and under the assumption of locally thermal populations, the late-time recovery of purity must follow a coherence-dominated route. We do this at two levels. First, in the simplest two-level setting, we construct explicit families of states with the same eigenvalues but opposite purity structure: one family with purity entirely in populations and another with purity entirely in coherences. This illustrates in the clearest possible way that global purity and spectrum do not uniquely determine how information is encoded. Second, we consider a many-qubit Page-type model in which the black hole and radiation form a bipartite pure state with varying Hilbert-space dimensions. Using known results on typical reduced states in large dimensions, we analyze how the indices $P_p$ and $P_c$ behave as the radiation subsystem grows and the black hole shrinks.

The analysis reveals a simple but important feature: in the large-dimension regime of a Page-type model, and if the radiation populations in the energy basis are constrained to remain close to thermal, the population purity index $P_p(k)$ remains small, while the coherence purity index $P_c(k)$ must grow and eventually account for almost all the global purity of the late-time radiation. In the geometric $(P_p,\tilde P_c)$ representation (with $\tilde P_c = \sqrt{\alpha_n}\,P_c$), the typical trajectory of the radiation state starts near the origin (maximally mixed), stays close to it around the Page time, and then moves towards the upper point of the unit circle $(P_p,\tilde P_c)\simeq (0,1)$, indicating a coherence-dominated route to purity.

Although our main technical analysis is performed within the standard, static Page model, we argue in the Discussion (Section~\ref{sec:discussion}) that the conclusion---the necessity of a coherence-dominated route under the assumption of locally thermal populations---is robust and expected to hold in more realistic, dynamical models of evaporation that exhibit quantum chaotic scrambling.

Summarizing our main conclusion, we show that in Page-type evaporation models with locally thermal radiation populations, the late-time recovery of purity is necessarily coherence-dominated: the population index remains small while the coherence index grows to carry essentially all the purity. In the $(P_p,\tilde P_c)$ plane this corresponds to trajectories ending near $(0,1)$, in contrast to population-dominated routes near $(1,0)$.

Our conclusions are explicitly derived within a Page-type setting without
an explicit Hamiltonian. We therefore regard the coherence-dominated
route identified here as a robust kinematic feature of unitary evaporation
under locally featureless populations, rather than as a complete dynamical
model of black-hole evaporation.

The paper is organized as follows. Section~\ref{sec:formalism} recalls the definition of the global purity index $P_{(n)}$ and the associated population and coherence indices $P_p$ and $P_c$ for generic density matrices, emphasizing their interpretation and the basic decomposition~\eqref{eq:Pn2_Pp_Pc_intro}. Section~\ref{sec:two_level} analyzes the two-dimensional case in detail, constructing explicit examples of population-dominated and coherence-dominated routes to purity and introducing the $(P_p,\tilde P_c)$ plane as a simple geometric representation. Section~\ref{sec:Page_model} applies these ideas to a standard Page-type evaporation model with $N$ qubits, derives the typical behavior of the radiation purity, and shows that, under locally thermal populations, the late-time purity must be carried by the coherence index $P_c$. Section~\ref{sec:discussion} concludes with a discussion of how this framework complements entropic analyses of the information problem, outlines possible extensions to more elaborate models of quantum gravity and to higher-order descriptors of statistical structure, and examines the robustness of our conclusions in dynamical evaporation scenarios.

\section{Global Purity and Population--Coherence Decomposition}
\label{sec:formalism}

In this section we recall the definition of the global purity index $P_{(n)}$ for an $n$-dimensional density matrix, and outline the associated decomposition into population and coherence contributions, $P_p$ and $P_c$. The detailed construction and normalization of these indices were originally developed in the context of polarization optics and general density-matrix analysis~\cite{GilSourcesAsymmetry}; here we present only the elements needed for the present application to black-hole radiation.

\subsection{Global purity index \texorpdfstring{$P_{(n)}$}{P(n)}}

Let $\rho$ be an $n\times n$ density matrix, i.e., a positive semidefinite, unit-trace Hermitian matrix acting on an $n$-dimensional Hilbert space. The standard purity
\begin{equation}
  \mathrm{Tr}(\rho^2)
\end{equation}
ranges from $1/n$ for the maximally mixed state $\rho = \mathbb{I}_n/n$ to $1$ for pure states $\rho = |\psi\rangle\langle\psi|$. A convenient normalized measure of global purity is given by~\cite{GilSourcesAsymmetry}
\begin{equation}
  P_{(n)}
  = \sqrt{\frac{n\,\mathrm{Tr}(\rho^2) - 1}{\,n-1\,}},
  \label{eq:Pn_def}
\end{equation}
which satisfies
\begin{equation}
  0 \le P_{(n)} \le 1,
\end{equation}
with
\begin{equation}
  P_{(n)} = 0 \ \Leftrightarrow\ \rho = \frac{\mathbb{I}_n}{n},
  \qquad
  P_{(n)} = 1 \ \Leftrightarrow\ \rho^2 = \rho.
\end{equation}

Equation~\eqref{eq:Pn_def} depends only on $\mathrm{Tr}(\rho^2)$, or equivalently on the quadratic invariant $\sum_j \lambda_j^2$ of the eigenvalues $\{\lambda_j\}$ of $\rho$. It is insensitive to the basis in which $\rho$ is expressed, but it does not distinguish whether purity arises from anisotropic level populations or from nontrivial off-diagonal structure in an appropriate intrinsic (IRB) decomposition.

\subsection{Separation of populations and coherences}

To separate population and coherence contributions, it is convenient to work
with an \emph{intrinsic density matrix} associated with $\rho$. In the
formalism developed in Ref.~\cite{GilSourcesAsymmetry}, one starts from the
decomposition
\begin{equation}
  \rho = \Re(\rho) + i\,\Im(\rho),
  \label{eq:rho_decomp}
\end{equation}
where $\Re(\rho)$ and $\Im(\rho)$ are Hermitian and $\Im(\rho)$ has real
antisymmetric matrix elements in any orthonormal basis. One then constructs a
real orthogonal matrix $Q$ that diagonalizes the real part $\Re(\rho)$. The
transformed density matrix
\begin{equation}
  \rho_O = Q\,\rho\,Q^\dagger
  \label{eq:rhoO_def}
\end{equation}
is said to be expressed in the \emph{intrinsic basis}, and $\rho_O$ itself
will be referred to as the intrinsic density matrix~\cite{GilSourcesAsymmetry}. In general, populations
are basis-dependent; what singles out the intrinsic basis is precisely that
in this representation the real part is diagonal, so that populations and
coherences are cleanly separated.

More explicitly, writing
\begin{equation}
  \rho_O = A + iN,
  \qquad
  A = \Re(\rho_O),
  \quad
  N = \Im(\rho_O),
  \label{eq:rhoO_A_N}
\end{equation}
one has that $A$ is a real diagonal matrix and $N$ is a real antisymmetric
matrix ($N^T=-N$, $n_{ii}=0$). The diagonal entries of $A$ define a set of
\emph{intrinsic populations} $\{a_k\}_{k=1}^n$, with $\sum_k a_k = 1$ and $A = \mathrm{diag}(a_1,\dots,a_n)$.

These intrinsic populations coincide with the eigenvalues of the real part
$\Re(\rho)$, and by convention they are ordered in nonincreasing order~\cite{GilSourcesAsymmetry}, $a_1 \ge a_2 \ge \dots \ge a_n,$ so as to fix unambiguously the intrinsic basis and the associated ordering.
In this frame the populations are thus encoded in the diagonal of $\rho_O$
and are decoupled from the coherences. Note that, when $\Re(\rho)$ has degenerate eigenvalues (e.g., in the two-level case $a_1=a_2=1/2$ so that $\Re(\rho)\propto \mathbb{I}$), the intrinsic basis is not unique: any real orthogonal rotation within the degenerate subspace leaves $\Re(\rho)$ diagonal. This non-uniqueness is harmless, because the intrinsic invariants $P_p$ and $P_c$ are basis-independent. In particular, for $n=2$ and $a_1=a_2$, one always has $P_p=0$, while $P_c$ remains well defined (e.g., $P_c=2|\Im(\rho_{01})|$.

Conversely, the off-diagonal elements of $\rho_O$ are purely imaginary:
$\Re(\rho_O)=A$ is diagonal and $\Im(\rho_O)=N$ has vanishing diagonal.
The imaginary off-diagonal entries represent the intrinsic coherences or
correlations of the state and are collectively parametrized by a set of real
intrinsic coherence parameters $\{n_{ij}\}$, $1\le i<j\le n$, through the
antisymmetric real matrix $N=(n_{ij})$. The transformation from the original basis
to the intrinsic basis is an orthogonal similarity transformation (a
particular case of a unitary transformation), so scalar quantities
constructed from $\{a_k\}$ and $\{n_{ij}\}$ are invariant.

Note that in Ref.~\cite{GilSourcesAsymmetry}, the intrinsic decomposition is written as
$\rho_O = A + iN_{\mathrm{(orig)}}/2$. Here we adopt the slightly different
convention~\eqref{eq:rhoO_A_N}, with $N = N_{\mathrm{(orig)}}/2$ absorbing
the factor $1/2$. The coherence index $P_c$ defined below is adjusted
accordingly so that its numerical value coincides with that in
Ref.~\cite{GilSourcesAsymmetry}; only the symbol used for the antisymmetric
matrix has changed.

Within this intrinsic representation, the global purity of $\rho$ can be
expressed in terms of a ``population part'' depending only on the
$\{a_k\}$ and a ``coherence part'' depending only on the matrix $N$.
Concretely, given the intrinsic density matrix $\rho_O$ with diagonal entries
$\{a_k\}_{k=1}^n$ and purely imaginary off-diagonal elements encoded in the
antisymmetric real matrix $N = (n_{ij})$, one introduces a
\emph{population purity index} $P_p$ as a normalized measure of the
anisotropy of the populations with respect to the equiprobable distribution
$1/n$. Explicitly~\cite{GilSourcesAsymmetry},
\begin{equation}
  P_p^2
  = \frac{n}{n-1}\sum_{k=1}^n\left(a_k-\frac{1}{n}\right)^2
  = \frac{n}{n-1}\left(\sum_{k=1}^n a_k^2 - \frac{1}{n}\right),
  \label{eq:Pp_def}
\end{equation}
so that $0\leq P_p\leq 1$ and
\begin{equation}
  P_p = 0 \quad\Leftrightarrow\quad a_k = \frac{1}{n}\ \ \forall k,
\end{equation}
i.e., $P_p$ depends only on the diagonal of $\rho_O$ and vanishes if and only
if all intrinsic populations are equal.

A second quantity is a \emph{coherence purity index} $P_c$, constructed from
the intrinsic off-diagonal quantities $\{n_{ij}\}$ and normalized so that it
measures the overall strength of intrinsic coherences. In terms of the  real antisymmetric matrix $N$ one can write
\begin{equation}
  P_c^2
  = 2\,\|N\|_F^2
  = 4\sum_{1\leq i<j\leq n} n_{ij}^2,
    \label{eq:Pc_def}
\end{equation}
where $\|\cdot\|_F$ denotes the Frobenius norm and the parameters $n_{ij}$ are real-valued by construction (since the corresponding off-diagonal entries of $\rho_O$ are $i\,n_{ij}$).
Again one has $0\leq P_c\leq 1$
and $P_c = 0 \Leftrightarrow\rho_O\ \text{is diagonal},$
while $P_c=1$ corresponds to the maximal allowed intrinsic coherence (or
correlation asymmetry) for a given dimension $n$~\cite{GilSourcesAsymmetry}.
In the two-dimensional case, Eq.~\eqref{eq:Pc_def} reduces to
$P_c = 2|\rho_{01}|$ in the intrinsic basis, in agreement with the standard
Bloch-sphere parametrization.

By construction, $P_c$ is the Frobenius-norm weight of the intrinsic antisymmetric imaginary part of $\rho_O$; equivalently, it quantifies how much of the global purity is stored in intrinsically off-diagonal structure (intrinsic coherence). In principle, $P_c$ can be accessed from full state tomography of $\rho$ (followed by the intrinsic-basis construction), or more economically via interferometric protocols that estimate off-diagonal correlators and purity-related quantities (e.g., two-copy interference/swap-type measurements of $\mathrm{Tr}(\rho^2)$ combined with diagonal constraints). While $P_c$ is determined by the reduced density matrix (a second-order object), scrambling diagnostics such as \emph{out-of-time-order correlators} (OTOCs) probe operator growth through higher-order (typically four-point) correlation functions. In many chaotic (strongly scrambling) unitary models, the rapid spreading of operators is accompanied by a fast build-up of nonlocal correlations. Under the additional assumption that the diagonal populations remain (approximately) thermal, such correlations are then expected to account for most of the purity recovery, which in our population--coherence decomposition manifests primarily as growth of $P_c$.

A short calculation using the intrinsic decomposition
$\rho_O = A + iN$ shows that
\begin{equation}
  \mathrm{Tr}(\rho^2)
  = \mathrm{Tr}(\rho_O^2)
  = \sum_{k=1}^n a_k^2 + \|N\|_F^2.
\end{equation}
Inserting Eqs.~\eqref{eq:Pp_def} and~\eqref{eq:Pc_def} into the definition
of the normalized purity $P_{(n)}$ in Eq.~\eqref{eq:Pn_def} then yields~\cite{GilSourcesAsymmetry}
\begin{equation}
  P_{(n)}^2 = P_p^2 + \alpha_n P_c^2,
  \qquad
  \alpha_n = \frac{n}{2(n-1)},
  \label{eq:Pn2_Pp_Pc_general}
\end{equation}
which is the general quadratic decomposition announced above. This relation
holds for any $n$-dimensional density matrix and, for $n=2$, reduces to the
simple Euclidean decomposition $P_{(2)}^2 = P_p^2 + P_c^2$ used in
Sec.~\ref{sec:two_level}.

It is important to emphasize that $P_p$ and $P_c$ are invariants of the
state, not quantities tied to a particular representation. They are defined
via the intrinsic decomposition \eqref{eq:rhoO_A_N}, but once the invariants
have been computed they can be associated with $\rho$ in any basis: changing
to another orthonormal basis simply acts by a unitary similarity
transformation and leaves $P_p$ and $P_c$ unchanged. In Sec.~\ref{sec:Page_model}
we will therefore be able to speak of population and coherence contributions
to purity in physically preferred bases (in particular, the energy basis of
the radiation) while still using the same intrinsic indices $P_p$ and $P_c$.

\subsection{Geometric representation in the \texorpdfstring{$(P_p,\tilde P_c)$}{(Pp, Pctilde)} plane}
\label{subsec:geo_plane}

For our purposes it is convenient to absorb the dimension-dependent factor $\alpha_n$ into a rescaled coherence index
\begin{equation}
  \tilde P_c \equiv \sqrt{\alpha_n}\,P_c
  = \sqrt{\frac{n}{2(n-1)}}\,P_c.
\end{equation}
In terms of $(P_p,\tilde P_c)$, the decomposition~\eqref{eq:Pn2_Pp_Pc_general} becomes
\begin{equation}
  P_{(n)}^2 = P_p^2 + \tilde P_c^2.
  \label{eq:Pn_circle}
\end{equation}

\begin{figure}[H]
\centering
\includegraphics[scale=0.5]{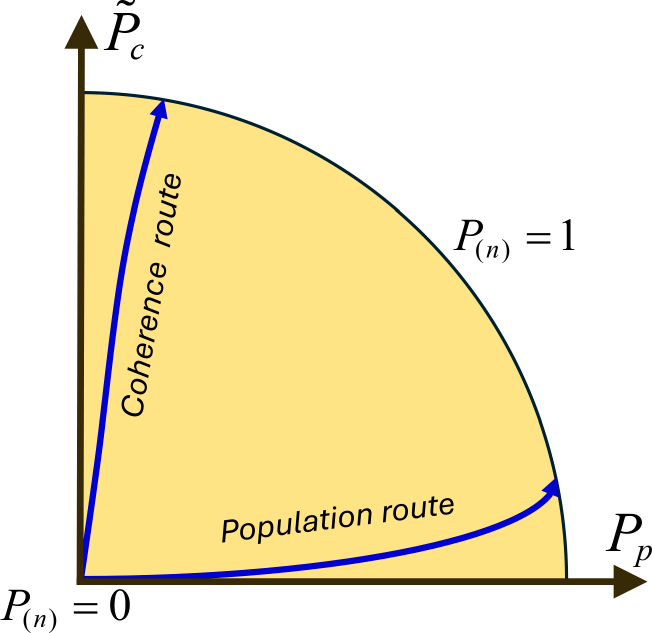}
\caption{
Population--coherence representation of density-matrix purity.
Each state is mapped to a point in the $(P_p,\tilde{P}_c)$ plane, with
$P_{(n)} = \sqrt{P_p^2 + \tilde{P}_c^2}$ as radial coordinate.
The origin corresponds to the maximally mixed state, while the unit circle
corresponds to pure states.
Horizontal trajectories illustrate population-dominated routes to purity,
and vertical trajectories illustrate coherence-dominated routes.
}
\label{fig:PopCohPlane}
\end{figure}

This suggests a natural geometric representation (see Figure~\ref{fig:PopCohPlane}). In this representation, each density matrix $\rho$ is associated with a point in the plane with coordinates $(P_p,\tilde P_c)$. The normalized global purity $P_{(n)}$ is the radial coordinate,
\begin{equation}
  P_{(n)} = \sqrt{P_p^2 + \tilde P_c^2},
\end{equation}
and the angle $\theta$ defined by
\begin{equation}
  P_p = P_{(n)}\cos\theta,\qquad
  \tilde P_c = P_{(n)}\sin\theta,
\end{equation}
quantifies the relative contribution of populations and coherences: $\theta \approx 0$ corresponds to purity dominated by populations, while $\theta \approx \frac{\pi}{2}$ corresponds to purity dominated by coherences.

In this picture, the origin $(P_p,\tilde P_c)=(0,0)$ corresponds to the maximally mixed state $P_{(n)}=0$, while the unit circle $P_{(n)}=1$ represents pure states. Points near $(1,0)$ describe states whose purity is almost entirely due to population asymmetry, whereas points near $(0,1)$ describe states whose purity is almost entirely due to coherence asymmetry. This population--coherence plane will play a central role in the rest of the paper.

\section{Two-Level Illustration: Population and Coherence Routes to Purity}
\label{sec:two_level}

In order to make the population--coherence decomposition fully transparent, it is useful to first analyze the simplest nontrivial case of a two-level system. In this section we show explicitly how the global purity can be split into population and coherence contributions, and we construct two opposite ``routes to purity'' that will later serve as prototypes for black-hole evaporation scenarios.

\subsection{Global, population, and coherence purity for \texorpdfstring{$n=2$}{n=2}}

Let $\rho$ be a $2\times2$ density matrix written in some fixed orthonormal basis
$\{\ket{0},\ket{1}\}$, which we shall regard as the energy basis for later physical
interpretation. We write
\begin{equation}
  \rho =
  \begin{pmatrix}
    \rho_{00} & \rho_{01} \\
    \rho_{10} & \rho_{11}
  \end{pmatrix},
  \qquad
  \rho_{10} = \rho_{01}^*,
  \qquad
  \rho_{00}+\rho_{11}=1.
\end{equation}
The standard purity satisfies
\begin{equation}
  \mathrm{Tr}(\rho^2)
  = \rho_{00}^2 + \rho_{11}^2 + 2|\rho_{01}|^2.
\end{equation}
For $n=2$ we introduce the normalized global purity index
\begin{equation}
  P \equiv P_{(2)}
  = \sqrt{2\,\mathrm{Tr}(\rho^2) - 1},
\end{equation}
which ranges from $P=0$ (maximally mixed state) to $P=1$ (pure state).

To define the population and coherence contributions consistently with the
general construction of Sec.~\ref{sec:formalism}, one must first pass to the
\emph{intrinsic} density matrix $\rho_O=Q\rho Q^\dagger$, where $Q$ is a real
orthogonal matrix that diagonalizes $\Re(\rho)$. In the intrinsic basis one
has $\rho_O=A+iN$, with $A=\mathrm{diag}(a_1,a_2),  (a_1 \ge a_2),$ and
$N=\bigl(\begin{smallmatrix}0&n\\-n&0\end{smallmatrix}\bigr)$, and the indices are
\begin{equation}
  P_p = |a_1-a_2|,
  \qquad
  P_c = 2|n| = 2\bigl|(\rho_O)_{01}\bigr|.
  \label{eq:PpPc_intrinsic_2D}
\end{equation}

In this $2\times2$ case, $P_p$ and $P_c$ can be expressed directly in terms of
the matrix elements of $\rho$ in the chosen basis. Writing
$\rho_{01}=x+iy$ with $x=\Re(\rho_{01})$ and $y=\Im(\rho_{01})$, the eigenvalues
of $\Re(\rho)=\bigl(\begin{smallmatrix}\rho_{00}&x\\ x&\rho_{11}\end{smallmatrix}\bigr)$ are
$a_{1,2}=\tfrac12\bigl[1\pm \sqrt{(\rho_{00}-\rho_{11})^2+4x^2}\bigr]$, so that
\begin{equation}
  P_p
  = \sqrt{(\rho_{00}-\rho_{11})^2 + 4\,\Re(\rho_{01})^2},
  \qquad
  P_c
  = 2\,|\Im(\rho_{01})|.
  \label{eq:PpPc_from_rho_2D}
\end{equation}
(Here we used that in two dimensions the antisymmetric part is invariant under
real orthogonal rotations, so the intrinsic parameter $n$ coincides with
$\Im(\rho_{01})$.)

In terms of the Bloch vector $\mathbf{r}=(r_x,r_y,r_z)$, with
\begin{equation}
  r_x = 2\Re(\rho_{01}),\quad
  r_y = -2\Im(\rho_{01}),\quad
  r_z = \rho_{00}-\rho_{11},
\end{equation}
we therefore have
\begin{equation}
  P = |\mathbf{r}|,\qquad
  P_p = \sqrt{r_x^2 + r_z^2},\qquad
  P_c = |r_y|.
\end{equation}
Thus, in the two-level case the decomposition takes the particularly simple form
\begin{equation}
  P^2 = P_p^2 + P_c^2,
  \label{eq:P2_Pp_Pc_2D}
\end{equation}
i.e., the global purity is the Euclidean norm of the two-component vector $(P_p,P_c)$.

Finally, note that if the chosen basis is already intrinsic (equivalently,
$\Re(\rho_{01})=0$, so the off-diagonal element is purely imaginary), then
Eq.~\eqref{eq:PpPc_from_rho_2D} reduces to the familiar simplified expressions
$P_p=|\rho_{00}-\rho_{11}|$ and $P_c=2|\rho_{01}|$.

In the $(P_p,P_c)$ plane, each state $\rho$ corresponds to a point with coordinates
\begin{equation}
  (P_p,P_c) = \bigl(\sqrt{(\rho_{00}-\rho_{11})^2 + 4\,\Re(\rho_{01})^2},\,2|\Im(\rho_{01})|\bigr),
\end{equation}
and fixed global purity $P$ corresponds to a circle of radius $P$. The maximally mixed state $\rho=\mathbb{I}/2$ lies at the origin $(0,0)$, whereas pure states lie on the unit circle $P=1$.

\subsection{Route A: purity stored in populations}

Our first example is a family of purely diagonal states of the form
\begin{equation}
  \rho_{\mathrm{diag}}(p)
  =
  \begin{pmatrix}
    1-p & 0 \\
    0 & p
  \end{pmatrix},
  \qquad 0 \le p \le 1,
  \label{eq:rho_diag}
\end{equation}
These states have eigenvalues $\lambda_1 = 1-p$ and $\lambda_2 = p$, and no coherence in the chosen basis:
\begin{equation}
  \rho_{01} = 0 \quad\Rightarrow\quad P_c = 0.
\end{equation}
The purity is
\begin{equation}
  \mathrm{Tr}\bigl(\rho_{\mathrm{diag}}^2\bigr)
  = (1-p)^2 + p^2
  = 1 - 2p + 2p^2,
\end{equation}
so the global purity index reads
\begin{equation}
  P_{\mathrm{diag}}(p)
  = \sqrt{2\bigl[(1-p)^2 + p^2\bigr] - 1}
  = |1 - 2p|.
\end{equation}
The population purity index is
\begin{equation}
  P_p(p) = |\rho_{00}-\rho_{11}| = |(1-p) - p| = |1-2p|,
\end{equation}
and therefore
\begin{equation}
  P_{\mathrm{diag}}(p) = P_p(p),\qquad P_c(p)=0,
\end{equation}
so that Eq.~\eqref{eq:P2_Pp_Pc_2D} is trivially satisfied.

This family provides a population-dominated route to purity. At $p=1/2$, one has $\rho_{\mathrm{diag}}(1/2)=\mathbb{I}/2$, so
\[
  P_{\mathrm{diag}}(1/2)=0,\quad P_p(1/2)=0,\quad P_c(1/2)=0,
\]
corresponding to the maximally mixed state at the origin $(P_p,P_c)=(0,0)$. When $p\to 0$ or $p\to 1$, the state becomes a pure basis vector, and
\[
  P_{\mathrm{diag}} \to 1,\quad P_p\to 1,\quad P_c\to 0,
\]
i.e., the point in the $(P_p,P_c)$ plane moves along the horizontal axis from the origin towards $(1,0)$.

Physically, this means that the global purity is entirely stored in the anisotropy of the level populations in the energy basis. If we interpret $\rho$ as the state of Hawking radiation in that basis, this scenario corresponds to a situation where information is recovered through strong deviations of the emission probabilities from a thermal spectrum.

\subsection{Route B: purity stored in coherences}

We now construct a second family of density matrices with the same eigenvalues as
$\rho_{\mathrm{diag}}(p)$, but for which the intrinsic populations are exactly
equiprobable, so that the global purity is carried entirely by intrinsic coherences.

Consider the unitary transformation
\begin{equation}
  U = \frac{1}{\sqrt{2}}
  \begin{pmatrix}
    1 & -i\\
    1 & \ \,i
  \end{pmatrix},
\end{equation}
and define
\begin{equation}
  \rho_{\mathrm{coh}}(p)
  = U\,\rho_{\mathrm{diag}}(p)\,U^\dagger.
\end{equation}
A simple calculation yields
\begin{equation}
  \rho_{\mathrm{coh}}(p)
  = \frac{1}{2}
  \begin{pmatrix}
    1 & i(1-2p) \\
    -i(1-2p) & 1
  \end{pmatrix}.
  \label{eq:rho_coh}
\end{equation}
One easily verifies that $\rho_{\mathrm{coh}}(p)$ has the same eigenvalues
$\lambda_{1,2}=1-p,p$ as $\rho_{\mathrm{diag}}(p)$, and hence the same global
purity
\begin{equation}
  P_{\mathrm{coh}}(p) = P_{\mathrm{diag}}(p) = |1-2p|.
\end{equation}

However, the internal organization of that purity is now completely different.
The real part of $\rho_{\mathrm{coh}}(p)$ is proportional to the identity,
\begin{equation}
  \Re\bigl(\rho_{\mathrm{coh}}(p)\bigr)=\frac{\mathbb{I}}{2},
\end{equation}
so the intrinsic populations are strictly equiprobable,
$a_1=a_2=\tfrac12$, and therefore
\begin{equation}
  P_p(p)=|a_1-a_2|=0.
\end{equation}
At the same time, the off-diagonal element is purely imaginary,
\begin{equation}
  \rho_{01} = \frac{i(1-2p)}{2}
  \quad\Rightarrow\quad
  P_c(p)=2|\Im(\rho_{01})|=|1-2p|.
\end{equation}
Therefore,
\begin{equation}
  P_{\mathrm{coh}}(p) = P_c(p),\qquad P_p(p)=0,
\end{equation}
so that the global purity is stored entirely in the intrinsic coherences.

In the $(P_p,P_c)$ plane, the trajectory of $\rho_{\mathrm{coh}}(p)$ is now as follows.
At $p=1/2$, one has $\rho_{\mathrm{coh}}(1/2) = \mathbb{I}/2$, so
\[
  P_p(1/2)=0,\quad P_c(1/2)=0,
\]
and the state starts at the origin. When $p\to 0$ or $p\to 1$, the state approaches
a pure state (with fixed equiprobable intrinsic populations), and
\[
  P_{\mathrm{coh}} \to 1,\quad P_c\to 1,\quad P_p\to 0,
\]
i.e., the point moves along the vertical axis towards $(0,1)$.

This family illustrates a coherence-dominated route to purity: the intrinsic populations remain strictly equiprobable, while the intrinsic coherences grow and carry all the purity.

\subsection{Physical interpretation}

The two families $\rho_{\mathrm{diag}}(p)$ and $\rho_{\mathrm{coh}}(p)$ demonstrate that even for fixed eigenvalues and fixed global purity $P$, the way in which information is encoded in the density matrix can be qualitatively different: in $\rho_{\mathrm{diag}}(p)$, the purity is entirely associated with population asymmetry in the energy basis ($P_p = P$, $P_c=0$), whereas in $\rho_{\mathrm{coh}}(p)$, the purity is entirely associated with coherences and correlations in that basis ($P_c = P$, $P_p=0$).

From the viewpoint of black-hole evaporation, if we model the late-time state of Hawking radiation as a mixed state close to thermal in its energy populations, then insisting that $P_p$ remains small effectively forbids a route of type $\rho_{\mathrm{diag}}(p)$. Any recovery of purity compatible with locally thermal populations must instead resemble a route of type $\rho_{\mathrm{coh}}(p)$, where the information is carried predominantly by coherences. In Section~\ref{sec:Page_model} we will see how this intuition is realised, in a statistical sense, in a many-qubit Page-type evaporation model.

\section{Page-Type Evaporation Model and Large-Dimension behavior}
\label{sec:Page_model}

We now consider a standard Page-type model of black-hole evaporation and analyze how the population and coherence purity indices behave in the large-dimension regime. This will show that, under the assumption of locally thermal populations in the energy basis, the late-time recovery of purity must be driven predominantly by the coherence contribution $P_c$.

\subsection{Page-type evaporation with \texorpdfstring{$N$}{N} qubits}

We model the black hole (BH) and its emitted radiation (rad) as a bipartite quantum system with Hilbert space
\begin{equation}
  \mathcal{H} = \mathcal{H}_{\rm BH}(k)\otimes \mathcal{H}_{\rm rad}(k),
\end{equation}
where $k=0,1,\dots,N$ counts the number of emitted qubits. Initially, the black hole consists of $N$ qubits and the radiation is empty. After the emission of $k$ qubits, the dimensions are
\begin{equation}
  d_{\rm BH}(k) = 2^{N-k},\qquad
  d_{\rm rad}(k) = 2^{k},
\end{equation}
so that the total dimension is conserved, $d_{\rm BH}(k)\,d_{\rm rad}(k)=2^{N}$.

Following Page's original approach~\cite{Page1993}, we assume that at each step the global state $\ket{\Psi_k} \in \mathcal{H}$ is a typical pure state drawn from the Haar measure on $\mathcal{H}$ compatible with the above bipartition. The Haar-random ensemble, while static, is known to accurately capture the \emph{average} behavior of reduced states under chaotic, scrambling dynamics~\cite{Hayden2007, Page1993}. It therefore serves as a reliable proxy for studying the kinematic constraints that any unitary evaporation model must satisfy, a point we revisit in Section~\ref{subsec:dynamics} when discussing dynamical extensions. The reduced state of the radiation is then
\begin{equation}
  \rho_{\rm rad}(k)
  = \operatorname{Tr}_{\rm BH} \bigl(\ket{\Psi_k}\bra{\Psi_k}\bigr),
\end{equation}
a mixed state on $\mathcal{H}_{\rm rad}(k)$ of dimension
\begin{equation}
  n(k) \equiv d_{\rm rad}(k) = 2^k.
\end{equation}

For any $n$-dimensional density matrix we define the normalized global purity index as in Eq.~\eqref{eq:Pn_def}. In the present context we write
\begin{equation}
  P(k)\equiv P_{(n(k))}(\rho_{\rm rad}(k))
\end{equation}
for the global purity of the radiation after $k$ emissions.

\subsection{Average purity of the radiation and Page-type behavior}

For a bipartite pure state on $\mathcal{H}_A\otimes\mathcal{H}_B$ with dimensions $d_A,d_B$, the average purity of the reduced state $\rho_A$ with respect to the Haar measure is known to be~\cite{Page1993}
\begin{equation}
  \bigl\langle \mathrm{Tr}(\rho_A^2)\bigr\rangle
  = \frac{d_A + d_B}{d_A d_B + 1}.
\end{equation}
Taking subsystem $A$ to be the radiation, we have $d_A=d_{\rm rad}(k)$, $d_B=d_{\rm BH}(k)$, so that
\begin{equation}
  \bigl\langle \mathrm{Tr}\bigl(\rho_{\rm rad}(k)^2\bigr)\bigr\rangle
  = \frac{d_{\rm rad}(k)+ d_{\rm BH}(k)}{d_{\rm rad}(k)\,d_{\rm BH}(k) + 1}.
\end{equation}
For large total dimension $2^N$, the ``+1'' in the denominator can be neglected, and we obtain the simple approximation
\begin{equation}
  \bigl\langle \mathrm{Tr}\bigl(\rho_{\rm rad}(k)^2\bigr)\bigr\rangle
  \approx \frac{1}{d_{\rm rad}(k)} + \frac{1}{d_{\rm BH}(k)}.
  \label{eq:Tr_rho2_Page}
\end{equation}

Substituting Eq.~\eqref{eq:Tr_rho2_Page} into Eq.~\eqref{eq:Pn_def} with $n(k)=d_{\rm rad}(k)$, we recover the typical Page-like behavior of the global purity $P(k)$. \textit{Early radiation} ($k\ll N/2$, $d_{\rm rad}(k)\ll d_{\rm BH}(k)$) corresponds to
\begin{equation}
  \mathrm{Tr}\bigl(\rho_{\rm rad}(k)^2\bigr) \approx \frac{1}{d_{\rm rad}(k)}
  \quad\Rightarrow\quad
  P(k)\approx 0,
\end{equation}
so the radiation is almost maximally mixed. 
This substitution is justified in the Page/Haar-typical setting: in high-dimensional bipartite systems, $\mathrm{Tr}(\rho_A^2)$ is sharply concentrated around its Haar average (concentration of measure). Therefore the typical purity curve is well approximated by $\langle \mathrm{Tr}(\rho_A^2)\rangle$, and using the mean provides an accurate analytic proxy for the behavior of almost all pure states in the ensemble (up to exponentially small deviations in the relevant dimensions).

\textit{Around the Page time} ($k\simeq N/2$, $d_{\rm rad}(k)\simeq d_{\rm BH}(k)$), one finds
\begin{equation}
  \mathrm{Tr}\bigl(\rho_{\rm rad}(k)^2\bigr) \approx \frac{2}{d_{\rm rad}(k)},
\end{equation}
and $P(k)$ reaches its minimum value, corresponding to maximal entanglement between black hole and radiation. \textit{Late radiation} ($k\lesssim N$, $d_{\rm rad}(k)\gg d_{\rm BH}(k)$) is characterized by
\begin{equation}
  \mathrm{Tr}\bigl(\rho_{\rm rad}(k)^2\bigr) \approx \frac{1}{d_{\rm BH}(k)} \gg \frac{1}{d_{\rm rad}(k)},
\end{equation}
and as the black-hole dimension approaches unity, $d_{\rm BH}(k)\to 1$, we have $\mathrm{Tr}(\rho_{\rm rad}(k)^2)\to 1$ and hence
\begin{equation}
  P(k)\to 1.
\end{equation}
The radiation becomes nearly pure as the black hole evaporates completely.

Thus, at the level of the global purity index $P(k)$, the model reproduces the characteristic Page curve: low purity at early times, a minimum around the Page time, and recovery of purity at late times. What is missing in this description is how this purity is encoded in the density matrix, i.e., whether it is stored in the diagonal populations or in the off-diagonal coherences. This is precisely what the indices $P_p$ and $P_c$ capture.

\subsection{Population and coherence indices in large dimension}

In the present application we are interested in populations in a physically
preferred basis for the radiation (ideally, the energy eigenbasis). The
diagonal entries in that basis are \emph{basis-dependent} and encode the local
spectrum seen by energy measurements. By contrast, the indices $P_p$ and $P_c$
are \emph{basis-invariant}: for each $k$ they are defined via the intrinsic
decomposition of $\rho_{\rm rad}(k)$, obtained by diagonalizing
$\Re\bigl(\rho_{\rm rad}(k)\bigr)$ with a real orthogonal matrix $Q(k)$ and
forming $\rho_{O}(k)=Q(k)\rho_{\rm rad}(k)Q(k)^\dagger=A(k)+iN(k)$. The indices
$P_p(k)$ and $P_c(k)$ then follow from Eqs.~\eqref{eq:Pp_def} and~\eqref{eq:Pc_def}.
When the preferred physical basis approximately diagonalizes
$\Re\bigl(\rho_{\rm rad}(k)\bigr)$ (i.e., the real off-diagonal part is
negligible), the intrinsic populations $\{a_i(k)\}$ are well approximated by the
physical populations $\{p_i(k)\}$, and the intrinsic coherences are directly
reflected in the off-diagonal structure in that basis.

We now fix a physically meaningful basis in $\mathcal{H}_{\rm rad}(k)$, which we identify with the energy eigenbasis of the emitted quanta. In that basis, the diagonal entries
\begin{equation}
  p_i(k) = \bigl(\rho_{\rm rad}(k)\bigr)_{ii},\qquad i=1,\dots,n(k),
\end{equation}
represent the level populations, while the off-diagonal elements
\begin{equation}
  \bigl(\rho_{\rm rad}(k)\bigr)_{ij},\qquad i\neq j,
\end{equation}
encode coherences and correlations between different energy levels.
These matrix elements are basis-dependent. Throughout the paper the indices $P_p$ and $P_c$ are defined from the intrinsic (IRB) form of $\rho_{\rm rad}(k)$ (Sec.~\ref{sec:formalism}), so the ``off-diagonal'' contribution relevant to $P_c$ should be understood in that intrinsic basis (and is directly visible in the energy basis only when $\Re\bigl(\rho_{\rm rad}(k)\bigr)$ is already approximately diagonal there).

As in Section~\ref{sec:formalism}, we compute for each $k$ the intrinsic indices $P_p(k)$ and $P_c(k)$ associated with $\rho_{\rm rad}(k)$, which satisfy
\begin{equation}
  P_{(n)}^2(k) = P_p^2(k) + \alpha_n P_c^2(k),
  \qquad
  \alpha_n=\frac{n}{2(n-1)}.
  \label{eq:P_Pp_Pc_general}
\end{equation}

In the random Page model, standard concentration-of-measure arguments imply that in the chosen energy basis the diagonal elements of $\rho_{\rm rad}(k)$ have average
\begin{equation}
  \bigl\langle p_i(k)\bigr\rangle = \frac{1}{n(k)},
\end{equation}
with typical fluctuations
\begin{equation}
  \delta p_i(k) \equiv p_i(k) - \frac{1}{n(k)}
  \sim \frac{1}{n(k)\sqrt{d_{\rm BH}(k)}}.
\end{equation}
Here and in the following, the symbol ``$\sim$'' is used to indicate
order-of-magnitude behavior in the limit of large $d_{\rm BH}(k)$, as
obtained from standard random-matrix (Wishart) estimates for typical
reduced states. The precise numerical factors are not important for our
conclusions and will not be needed.

Standard random-matrix results for Wishart ensembles imply that, in the limit
of large $d_{\rm BH}(k)$, these fluctuations are of order
$1/[n(k)\sqrt{d_{\rm BH}(k)}]$.

When the black-hole dimension is large, $d_{\rm BH}(k)\gg 1$, these fluctuations are extremely small. Moreover, the typical \emph{real} off-diagonal entries of $\rho_{\rm rad}(k)$ are of the same order as $\delta p_i(k)$, so the symmetric matrix $\Re\bigl(\rho_{\rm rad}(k)\bigr)$ is close to $\mathbb{I}/n(k)$ up to a small random perturbation. Consequently, its eigenvalues (the intrinsic populations $a_i(k)$) remain close to $1/n(k)$, and one expects
\begin{equation}
  P_p(k) \approx 0,
  \label{eq:Pp_small}
\end{equation}
consistent with locally featureless (thermal-like) populations in the chosen physical basis. The off-diagonal elements satisfy
\begin{equation}
  \bigl\langle \bigl|\bigl(\rho_{\rm rad}(k)\bigr)_{ij}\bigr|^2\bigr\rangle
  \sim \frac{1}{n(k)^2\,d_{\rm BH}(k)},
  \qquad i\neq j,
\end{equation}
so each individual coherence is small, but there are $n(k)^2-n(k)$ such elements. The total contribution of the off-diagonal terms to the purity $\mathrm{Tr}(\rho_{\rm rad}^2)$ then scales as
\begin{equation}
  \sum_{i\neq j} \bigl|\bigl(\rho_{\rm rad}(k)\bigr)_{ij}\bigr|^2
  \sim n(k)^2 \cdot \frac{1}{n(k)^2\,d_{\rm BH}(k)}
  = \frac{1}{d_{\rm BH}(k)},
\end{equation}
i.e., with the same scaling as the late-time contribution in Eq.~\eqref{eq:Tr_rho2_Page}.
These scalings follow from the same Wishart-type considerations, now applied
to the off-diagonal entries, and are again to be understood as asymptotic
estimates for large $d_{\rm BH}(k)$.

Thus, in the late radiation regime $d_{\rm rad}(k)\gg d_{\rm BH}(k)$, the dominant contribution to the global purity originates from the off-diagonal structure of $\rho_{\rm rad}(k)$. If, in addition, we impose the physically motivated condition that, in this simplified Page model without explicit energy levels, the populations remain nearly featureless and close to the uniform distribution in the chosen basis---so that $P_p(k)$ stays small throughout the evaporation---then the decomposition~\eqref{eq:P_Pp_Pc_general} forces the coherence index $P_c(k)$ to grow and eventually carry almost all the purity:

\begin{equation}
  P_{(n)}^2(k_{\rm final}) \approx \alpha_n\,P_c^2(k_{\rm final}),\qquad P_p(k_{\rm final})\approx 0,
\end{equation}
and hence $P_c(k_{\rm final})$ must be close to its maximal value.

In other words, for a typical Page-type evolution with locally thermal populations, the recovery of global purity at late times cannot be attributed to a strong population asymmetry. Instead, it must be encoded in a build-up of coherence asymmetry, quantified by $P_c(k)$.

\subsection{Trajectories in the \texorpdfstring{$(P_p,\tilde P_c)$}{(Pp, Pctilde)} plane}
\label{subsec:page_traj_plane}

We use the population--coherence plane introduced in Sec.~\ref{subsec:geo_plane} to visualize the evolution of the radiation state. At each emission step $k$ we plot the point $(P_p(k),\tilde P_c(k))$, with
\begin{equation}
  \tilde P_c(k) \equiv \sqrt{\alpha_{n(k)}}\,P_c(k),
  \qquad n(k)=2^k.
\end{equation}
By the decomposition~\eqref{eq:P_Pp_Pc_general} (equivalently, Eq.~\eqref{eq:Pn_circle} applied at each $k$), the radial coordinate is the global purity $P_{(n(k))}(k)$.

In this representation, early times (and around the Page time) correspond to points close to the origin. In the late-time regime, if the radiation populations remain locally featureless so that $P_p(k)$ stays small, the recovery of global purity $P_{(n(k))}(k)\to 1$ forces $\tilde P_c(k)\to 1$. Geometrically, the typical trajectory therefore approaches the upper point of the unit circle, $(P_p,\tilde P_c)\simeq (0,1)$, i.e., a coherence-dominated route to purity (the many-body analogue of Route~B in Sec.~\ref{sec:two_level}).

\section{Discussion}
\label{sec:discussion}

The analysis presented in this work suggests that the population--coherence decomposition of purity provides a useful complementary perspective on the black-hole information problem. Rather than focusing solely on entropic quantities, we have examined how the global purity of the Hawking radiation can be split into contributions from level populations and coherences in a physically meaningful basis, and how these contributions evolve in simple models of evaporation.

At the most elementary level, the two-level examples of Section~\ref{sec:two_level} already capture an important conceptual point: even when the eigenvalues of a density matrix, and thus its entropy and global purity, are fixed, the internal structure of the density matrix can differ substantially. One family of states can realise purity almost entirely through population asymmetry ($P_p \simeq P$, $P_c \simeq 0$), while another family with the same eigenvalues can realise the same purity through coherences ($P_c \simeq P$, $P_p \simeq 0$). In the $(P_p,\tilde P_c)$ representation, these correspond to orthogonal routes in the population--coherence plane. This demonstrates that knowing how pure the radiation is does not tell us how the information is encoded---whether in emission probabilities or in correlations.

The Page-type model of Section~\ref{sec:Page_model} extends this insight to a many-body setting that captures the essential kinematics of black-hole evaporation. In such a model, the black hole and radiation form a bipartite pure state with varying Hilbert-space dimensions, and the average purity of the radiation follows a Page-like behavior: nearly zero at early times, minimal around the Page time, and approaching unity as the black hole evaporates. When we supplement this picture with the requirement that the radiation remain locally thermal in its energy populations, the population purity index $P_p(k)$ is constrained to remain small throughout the evolution. Under these conditions, the late-time recovery of global purity $P_{(n)}(k)\to 1$ can only be achieved via an increase of the coherence index $P_c(k)$. In the $(P_p(k),\tilde P_c(k))$ plane, with
$\tilde P_c(k) \equiv \sqrt{\alpha_{n(k)}}\,P_c(k)$ as introduced in
Sec.~\ref{sec:formalism}, this corresponds to a trajectory that starts at the origin and ends near $(P_p,\tilde P_c)\simeq (0,1)$: a coherence-dominated route to purity.

This observation does not, by itself, solve the information problem. It does, however, make more precise a qualitative idea that is often invoked in discussions of unitary evaporation: that the late-time radiation must carry information in subtle correlations rather than in its local spectrum. The indices $P_p$ and $P_c$ provide a way to quantify this statement. Population-dominated scenarios, in which information reappears primarily through large deviations from a thermal distribution of emission probabilities, would correspond to trajectories that end near $(P_p,\tilde P_c)\simeq (1,0)$, and are therefore incompatible with the assumption of locally thermal populations. Coherence-dominated scenarios, in which the spectrum remains nearly thermal while purity is restored, correspond instead to trajectories approaching $(0,1)$, as in the Page-type model with locally thermal diagonals.

From this viewpoint, the population--coherence plane can be regarded as a kind of phase diagram of information recovery routes. Different proposals for resolving the information problem---whether they rely on modified near-horizon physics, nonlocal interactions, or quantum-gravity-induced correlations---can, in principle, be mapped to different regions and trajectories in this plane, depending on how they partition purity between $P_p$ and $P_c$. This suggests that the pair $(P_p,P_c)$ may serve as a useful diagnostic for comparing and classifying evaporation models, much as entanglement entropies and Page curves are used today.

\subsection{Robustness beyond the static Page model: Dynamics, scrambling, and toy models}
\label{subsec:dynamics}

The analysis of Section 4 relied on the canonical Page model, in which the global state at each step is drawn from the Haar-random ensemble. While this approach correctly captures the kinematic constraints of unitary evolution in a bipartite system with changing dimensions, it does not prescribe a specific dynamical law for the emission process. A natural question is therefore whether the coherence-dominated route to purity is an artifact of this equilibrium ensemble, or a robust outcome of more realistic, time-dependent evaporation dynamics.

A growing body of work indicates that the key features of the Page model---in particular, the Page curve for entanglement entropy---emerge naturally from \emph{chaotic unitary dynamics} that scrambles information efficiently. In models of black-hole evaporation based on random unitary circuits~\cite{Hayden2007, Sekino2008, Hosur2016} or on fast-scrambling Hamiltonians~\cite{Shenker2013}, the reduced state of the radiation at a given time is statistically indistinguishable from that of a Haar-random state, provided the dynamics is sufficiently scrambling and the black hole has thermalized internally. The underlying mechanism is that rapid scrambling effectively ``resets'' the black-hole interior to a typical state after each emission, thereby justifying the use of random ensembles for time-averaged properties. Consequently, the statistical arguments leading to Eqs.~(53) and (55)---the strong concentration of diagonal populations and the collective buildup of off-diagonal contributions---are expected to hold in any chaotic dynamical model that reproduces the Page curve. Under the same physical requirement of locally thermal populations, the late-time purity must therefore still be carried predominantly by $P_c$.

This expectation is further supported by explicit toy models of quantum gravity in low dimensions. In Jackiw--Teitelboim (JT) gravity coupled to matter, the evaporation process can be described microscopically by a deterministic (though complex) unitary evolution of a boundary theory~\cite{Almheiri2019, Almheiri2020}. The celebrated island formula, which yields the Page curve for entanglement entropy, emerges from a gravitational path integral that sums over spacetime geometries with replica wormholes. Although a direct computation of $P_p$ and $P_c$ in such models is technically challenging, the gravitational prescription implies that the radiation density matrix, in the regime where a quantum extremal surface (island) dominates, is well approximated by a \emph{thermal density matrix corrected by entanglement across the wormhole}~\cite{Almheiri2019,Almheiri2020,Penington2019}, schematically the structure of a state with small $P_p$ but substantial $P_c$, for example via replica wormholes inducing
correlations between radiation and interior degrees of freedom that, for the reduced radiation state, manifest as intrinsic coherences
(a nonzero antisymmetric part $N$ in the IRB form of $\rho_{\rm rad}$).
In the language developed in this paper, such correlations can be viewed as precisely the kind of intrinsic off-diagonal
(purely imaginary in the IRB) structure that increases the coherence index $P_c$ while leaving the population index $P_p$ nearly
unchanged. This provides suggestive evidence that the coherence-dominated
route is not a mere artifact of random averaging, but a generic consequence
of unitary evaporation that is consistent with semiclassical gravity, and
that semiclassical island formulas and replica-wormhole corrections offer
concrete realizations of such coherence-dominated routes in gravitational
settings.

It is instructive, however, to consider scenarios where the coherence-dominated route might be modified. In systems with weak scrambling, integrable dynamics, or symmetries that protect population anisotropies, the trajectory in the $(P_p,\tilde P_c)$ plane could deviate significantly from the vertical path. For example, in a free (non-interacting) emission model, information might escape through systematic deviations in the energy spectrum, leading to a more horizontal trajectory. The population--coherence plane can thus serve as a diagnostic tool to classify evaporation models according to their scrambling strength and dynamical properties. A concrete numerical study of $(P_p(k),P_c(k))$ in random unitary circuit models---where the evolution can be followed step by step---would be a valuable future direction to quantify these deviations and to identify regimes where population effects become non-negligible.

The coherence-dominated conclusion hinges on restricting the \emph{diagonal} structure of the radiation in a physically preferred basis (typically the energy basis) to remain close to thermal (or, more generally, close to locally featureless/equiprobable populations in the intrinsic sense). The motivation is semiclassical: Hawking’s calculation predicts that each outgoing mode is approximately thermal when coarse-grained, and many unitary toy models of evaporation reproduce Page-curve behavior while keeping local spectra close to thermal as a consequence of efficient scrambling and conservation laws. In our language, this is precisely the regime where $P_p(k)$ stays small.

If local radiation is not thermal, the formalism and decomposition remain valid, but the endpoint in the $(P_p,\tilde P_c)$ plane need not approach $(0,1)$. Instead, deviations from thermality allow part of the recovered global purity to be carried by population anisotropy, i.e., trajectories can interpolate between coherence-dominated and population-dominated routes. Thus, the present framework makes transparent \emph{how much} purity recovery can be attributed to non-thermal diagonals (via $P_p$) versus genuinely off-diagonal correlations (via $P_c$).

In summary, while the static Page model offers the clearest analytic window, the physical heart of the argument---that a large, chaotic environment (the black hole) enforces locally thermal diagonals, while unitarity forces global purity to be recovered via coherences---is inherently dynamical. The population--coherence decomposition therefore provides a refined lens to investigate not only \emph{whether} information returns in a given model of quantum gravity, but \emph{how} it returns.

\subsection{Thermal reference states and energy-resolved extensions}
\label{subsec:thermal_reference}

In the analysis above, the Page model has been treated as an ``energy-free'' kinematic framework: the Hilbert space of the radiation is a tensor product of qubit spaces, but no explicit Hamiltonian or energy spectrum has been introduced. In this setting, the natural reference state for the diagonal of $\rho_{\rm rad}(k)$ is the uniform distribution, and small values of $P_p(k)$ simply express that the populations are close to equiprobable in the chosen basis. This is sufficient for our purposes here, since the argument only requires that the diagonal be nearly featureless, so that the recovery of purity must proceed via the intrinsic off-diagonal (purely imaginary in the IRB) structure quantified by $P_c(k)$.

In more realistic models of black-hole radiation, however, the Hilbert space comes equipped with a Hamiltonian $H$ and an associated thermal (Gibbs) state
\begin{equation}
  \rho_{\rm th} = \frac{e^{-\beta H}}{Z},
  \qquad
  Z = \mathrm{Tr}\bigl(e^{-\beta H}\bigr),
\end{equation}
which is diagonal in the energy eigenbasis with non-uniform populations $p_i^{(\mathrm{th})} \propto e^{-\beta E_i}$. With the convention adopted in this paper, $P_p$ measures deviations from the uniform distribution $1/n$, so a genuine thermal state at finite temperature has $P_p>0$ and is, in this sense, ``pure by populations''. Physically, however, what matters in such energy-resolved models is not the deviation from uniformity, but the deviation from the thermal reference distribution $p_i^{(\mathrm{th})}$.

This suggests a natural extension of the present framework in which the role of benchmark state is played by a fixed reference density matrix $\sigma$, typically chosen as $\sigma=\rho_{\rm th}$ or as a microcanonical state on a suitable energy window. One can then consider the difference
\begin{equation}
  \Delta \rho = \rho - \sigma,
\end{equation}
and decompose its structure into population and coherence parts in the energy basis: the diagonal of $\Delta\rho$ measures deviations of the actual populations from those of the reference state, while the off-diagonal entries capture coherences between different energy levels. In such a setting it would be natural to define ``thermal'' population and coherence indices, say $P_p^{(\mathrm{th})}$ and $P_c^{(\mathrm{th})}$, which quantify, respectively, the population anisotropy relative to $\rho_{\rm th}$ and the amount of coherence in the energy basis. These ``thermal'' indices are distinct from the intrinsic indices $P_p,P_c$ defined via the IRB rotation (Sec.~\ref{sec:formalism}); they coincide only when the energy basis approximately diagonalizes $\Re(\rho)$.

A fully developed version of this idea would involve choosing an appropriate normalization for $(P_p^{(\mathrm{th})},P_c^{(\mathrm{th})})$ and, ideally, constructing a relative purity measure that decomposes as
\begin{equation}
  P_{\rm rel}^2 = \bigl(P_p^{(\mathrm{th})}\bigr)^2 + \bigl(\tilde P_c^{(\mathrm{th})}\bigr)^2,
\end{equation}
with $\tilde P_c^{(\mathrm{th})}$ a suitably rescaled coherence index. In such a scheme, a strictly thermal state would be characterized by $P_p^{(\mathrm{th})}=0$ and $P_c^{(\mathrm{th})}=0$, while deviations from thermal behavior would be encoded in non-zero values of these indices.

The present work has deliberately focused on the simpler energy-free Page model, for which the uniform state plays the role of reference and the original indices $(P_p,P_c)$ suffice to distinguish population- and coherence-dominated routes to purity. A detailed development of thermal reference indices in energy-resolved evaporation models---including explicit Hamiltonians, greybody factors, and frequency-resolved radiation---is left for future work. Such an extension would provide a more realistic setting in which to test whether the coherence-dominated route identified here persists when the full energy structure of the Hawking spectrum is taken into account.

An important open technical aspect in this direction is the choice of
normalization for $P_p^{(\mathrm{th})}$ and $P_c^{(\mathrm{th})}$.
Unlike the uniform reference case, where the maximal values of the
indices are fixed by the dimension alone, a non-uniform thermal reference
$\rho_{\rm th}$ introduces an additional scale set by its spectrum.
Ensuring that the resulting thermal indices remain bounded (for instance,
between 0 and 1) and admit a clean decomposition of a suitably defined
relative purity will require a careful analysis of the allowed deviations
from $\rho_{\rm th}$. We leave this normalization problem, together with
explicit model studies, for future work.

\subsection{Connection with island formulas and entanglement wedges}

Modern treatments of the information problem express Page-curve behavior in terms of quantum extremal surfaces and islands, building on the quantum extremal surface prescription of Engelhardt and Wall~\cite{EngelhardtWallQES} and its applications to evaporating black holes and entanglement wedge reconstruction~\cite{Almheiri2019,Almheiri2020,Penington2019}. It would be interesting to investigate whether, in models where island calculations can be performed, the microscopic reduced states of the radiation (or suitable coarse-grained versions thereof) exhibit the population--coherence structures described here. In particular, one could ask whether the onset of an island corresponds to a qualitative change in the coherence index $P_c$, or in the angular parameter $\theta$ in the $(P_p,\tilde P_c)$ plane.

In modern gravitational computations of the Page curve, the restoration of unitarity is often attributed to nontrivial saddle points (replica wormholes) and the emergence of islands in the entanglement wedge. From the present viewpoint, such mechanisms can be interpreted as generating \emph{nonlocal correlations} that purify the radiation while leaving its coarse-grained local spectrum approximately thermal. This is naturally captured as a growth of the intrinsic coherence sector quantified by $P_c(k)$. Likewise, in random unitary circuit models with conservation laws, the local reduced states can remain close to thermal while purification proceeds through increasingly structured many-body correlations, again corresponding to a coherence-dominated trajectory in the $(P_p,\tilde P_c)$ plane.

For the definition and operational meaning of $P_c$---including its interpretation as a second-order (density-matrix) diagnostic and its distinction from higher-order scrambling probes such as out-of-time-order correlators (OTOCs)---we refer the reader to Sec.~2.2.

Recent laboratory analogs of horizon physics may offer settings where coherence structure is experimentally more accessible than in astrophysical situations. For instance, phase-space horizons in surface-gravity water waves and shallow-water analog black holes have been demonstrated in Refs.~\cite{Rozenman2024,Correa2025}. It would be interesting to explore whether a population--coherence decomposition of reconstructed density matrices (or mode covariance matrices) can help diagnose where information-like signatures reside in such analog systems.

\subsection{Multipartite structure and early/late radiation}

In the present work, the radiation has been treated as a single subsystem. A more refined analysis would partition it into early and late radiation, or into angular or frequency sectors, and study how purity and coherence are distributed among these sub-blocks. This could help clarify how information is shared between early and late Hawking quanta and whether certain sectors are particularly responsible for carrying coherence-dominated purity.

\subsection{Higher-order statistical descriptors}

The index $P_{(n)}$ is a second-order quantity, depending on $\mathrm{Tr}(\rho^2)$. In previous work on polarization and statistical optics, higher-order descriptors (e.g., fourth-order moments or combined indices such as $(\mu^{(2)},\mu^{(4)},S)$) have proven useful for refining the characterization of fluctuations and non-Gaussianity. Extending the population--coherence decomposition to such higher-order descriptors could provide additional insight into the fine structure of the radiation state, and perhaps distinguish between different coherence-dominated scenarios that look similar at second order.

\subsection{No-go statements and constraints}

Finally, the decomposition $P_{(n)}^2 = P_p^2 + \alpha_n P_c^2$ suggests the possibility of formulating simple inequalities that any unitary evaporation model must satisfy. For instance, if one imposes that $P_p(k)\le \varepsilon$ for all $k$ (a quantitative expression of ``almost thermal populations''), then any late-time purity $P_{(n)}(k_{\rm final})$ close to unity implies a lower bound on $P_c(k_{\rm final})$. Exploring such constraints in concrete models could lead to no-go statements for classes of semiclassical approximations that fail to generate sufficient coherence asymmetry. 

For instance, if for some $\varepsilon \ll 1$ one enforces $P_p(k)\le \varepsilon$ at all stages of evaporation, then for any final purity $P_{(n)}(k_{\rm final})\ge 1-\delta$ one must have
\begin{equation}
  P_c(k_{\rm final})
  \ge \frac{\sqrt{(1-\delta)^2 - \varepsilon^2}}{\sqrt{\alpha_n}},
\end{equation}
whenever $(1-\delta)^2 > \varepsilon^2$, so that the right-hand side is real and positive. Even such simple constraints already quantify the minimal coherence asymmetry required for a given level of information recovery under nearly thermal populations.

In summary, the population--coherence decomposition does not replace entropic methods or island calculations, but rather complements them by providing a spectral and structural refinement of what it means for the radiation to ``become pure again''. It exposes the internal organisation of purity within the density matrix and offers a compact geometric language---the $(P_p,\tilde P_c)$ plane---to describe different routes by which information may return. We hope that this framework will prove useful as a diagnostic tool in future studies of black-hole evaporation and, more generally, in the analysis of complex quantum systems where the distinction between population- and coherence-based information is physically meaningful.


\bigskip

\noindent\textbf{Funding.} No funding.

\smallskip
\noindent\textbf{Disclosures.} The author declares no conflicts of interest.

\smallskip
\noindent\textbf{Data availability.} No additional data were generated or analyzed in the course of this research.


\end{document}